\documentclass[a4paper,12pt]{article}
\pdfoutput=1 

\usepackage{jcappub} 
\usepackage{color}

\graphicspath{{figures/}}

\newcommand{\be}{\begin{equation}}
\newcommand{\ee}{\end{equation}}
\newcommand{\bea}{\begin{eqnarray}}
\newcommand{\eea}{\end{eqnarray}}
\newcommand{\beal}{\begin{aligned}}
\newcommand{\eeal}{\end{aligned}}

\newcommand{\Veff}{\ensuremath{V_{\text{eff}}}}

\title{Black hole accretion discs and screened scalar hair}

\author[a]{Anne-Christine Davis}
\author[b,c]{Ruth Gregory}
\author[a]{Rahul Jha}

\affiliation[a]{Department of Applied Mathematics and Theoretical Physics,
Centre for Mathematical Sciences, University of Cambridge, 
Wilberforce Road, Cambridge, CB3 0WA, U.K.}
\affiliation[b]{Centre for Particle Theory, Durham University,
South Road, Durham, DH1 3LE, UK}
\affiliation[c]{Perimeter Institute, 31 Caroline Street North, Waterloo, 
ON, N2L 2Y5,
Canada}

\emailAdd{acd@damtp.cam.ac.uk}
\emailAdd{r.a.w.gregory@durham.ac.uk}
\emailAdd{r.jha@damtp.cam.ac.uk}

\abstract{We present a novel way to investigate scalar field profiles around black 
holes with an accretion disc for a range of models where the Compton wavelength 
of the scalar is large compared to other length scales. By analysing the problem 
in ``Weyl" coordinates, we are able to calculate the scalar profiles for accretion
discs in the static Schwarzschild, as well as rotating Kerr, black holes. We 
comment on observational effects.}

\rightline{DCPT-16/25}

\begin{document} 
\maketitle
\flushbottom

\section{Introduction}\label{intro}

In modern cosmology, we are well aware we have to find a compelling
explanation for the late time acceleration of the universe -- one that fits
not only observation 
\cite{Perlmutter:1998np,Riess1998,Spergel:2003cb,Abazajian:2008wr,
Hinshaw:2012aka,Ade:2013zuv}, 
but also is theoretically consistent. One intriguing possibility is that of a 
light rolling scalar field. 
From a purely theoretical point of view, massless scalar fields or 
moduli are abundant in string and supergravity theories: Generic 
string compactifications result in a plethora of massless scalars 
in the low energy $4D$ effective theory, including Kaluza-Klein 
scalars describing the size of the compactified dimension; 
distances between branes in brane-world type scenarios
appear as scalar fields, and in addition to this, any supergravity theory 
requires scalar counterparts to all fermionic degrees of freedom thus
$\mathcal{N}\geq4$ supergravity  necessarily contains scalars 
in the gravity multiplets.

All of the above seems to suggest that adding scalars to the 
low energy description of gravity might be a reasonable thing to do. 
However, a famous theorem due to Weinberg 
\cite{Weinberg:1988cp} shows that any such modification necessarily 
introduces a new dynamical degree of freedom which in turn 
produces a fifth force. If the mediator of this force is light (which is 
necessary for the field to be of cosmological relevance) it would 
lead to unacceptably large violations of the Equivalence Principle 
(EP) within the solar system.Therefore if the reason for the late 
time acceleration of the universe is a scalar, there must be a 
mechanism to screen out its EP-violating effects. 

To explore such screening mechanisms, one schematically 
writes the scalar Lagrangian as
\begin{equation}
{\cal L} \supset \frac{Z(\phi_o)}{2}(\partial\delta\phi)^2
-\frac{m^2(\phi_o)}{2}(\delta\phi)^2+
\frac{\gamma(\phi_o)}{M_{\text{Pl}}}\delta\phi T
\end{equation}
where small variations of the scalar $\delta\phi$ around the 
background value $\phi_o$ can couple to the trace of the energy 
momentum tensor $T$. We can now see qualitatively the various
different screening scenarios: The scalar force is screened 
by the Vainshtein mechanism when $Z(\phi_o)$ is large enough 
that the canonically normalised field coupling,  
$\gamma(\phi_o)/Z^{1/2}(\phi_o)$, is small. The chameleon 
mechanism \cite{Khoury:2003rn,Brax:2004qh} occurs when the mass $m(\phi_o)$ 
is large enough to suppress the range of the scalar force. The 
symmetron \cite{Hinterbichler:2010es} and dilaton \cite{Brax:2010gi} screenings 
work by suppressing the scalar 
coupling $\gamma(\phi_o)$. In all of these cases, the 
background value of the field $\phi_o$ depends on the 
environment and the screening mechanisms occur in the 
presence of dense matter.

Obviously these theories have been scrutinised heavily in the laboratory
\cite{Brax:2007ak,Brax:2007vm,Upadhye2012b,Upadhye2012,
Burrage:2014oza,Elder:2016yxm}, 
solar system~\cite{Bertotti2003}, astrophysical~\cite{Burrage:2008ii,Jain2012},
and cosmological~\cite{Davis:2009vk,Jain2010} settings 
(see also \cite{Erickcek:2013oma} and \cite{Padilla:2015wlv} for interesting 
recent commentary on
the cosmological chameleon). All of these investigations, 
however, probe gravity in a regime where the gravitational fields 
and space-time curvatures are relatively weak. After the direct 
detection of gravitational waves from LIGO \cite{Abbott:2016blz}, we can now
hope that gravitational waves from compact binary systems will 
allow us to constrain the behaviour of gravity in the strong field, large 
curvature regime. Accordingly, attention has increasingly focussed on 
efforts to test gravity by studying the dynamics of compact objects such 
as neutron stars and black holes, \cite{Yunes2013,Psaltis2008}, and thus
it is natural to ask whether observations of black holes might provide new 
constraints on screened modified gravity 
\cite{Yunes2012,Mirshekari2013,Healy2011}. 

The constraints from pulsar systems \cite{Brax:2013uh} rely on the fact that in 
scalar-tensor theories neutron stars take on a non-constant scalar 
profile, producing equivalence principle violations. In the context of 
black holes and additional scalar degrees of freedom, however, the 
uniqueness of exact solutions needs to be carefully examined due a 
number of ``no-hair'' theorems that require the scalar fields to take on 
a constant value around isolated black holes
\cite{Bekenstein1996,Sotiriou2012,Hui:2012qt,Faraoni2013}.
This might seem to imply that black hole systems will 
not be useful for constraining screened modified gravity.  
However all of these no-hair theorems generally apply to black 
hole systems that are asymptotically flat, stationary, and include 
no matter -- hardly the typical galactic environment!  
By systematically relaxing these assumptions we can gain insight into 
scenarios where screened modified gravity may 
have non-trivial effects on black hole dynamics.  

Indeed, even without modifying gravity, there are many interesting
phenomena with scalars and non-static black holes. For example,
unstable massive scalar modes around rotating black holes 
\cite{Cardoso2005,Dolan2007,Cardoso2011,Dolan2012,Witek2013,Cardoso2013a},
or scalar hair around rotating black holes in Einstein gravity
\cite{Herdeiro:2014goa,Herdeiro:2015tia,Herdeiro:2015waa}.
It has also been shown that scalar hair will be induced if the asymptotic 
boundary conditions for the scalar field vary slowly with time
\cite{Jacobson1999,Frolov:2002va,Horbatsch2011,Chadburn:2013mta}
(see also \cite{Guo:2013dha} for a study of spherical collapse in scalar-tensor
gravity).  This time variation, which violates the conditions of stationarity 
and asymptotic flatness, could be due to either the cosmological 
evolution of the scalar field's background value or to the motion of the 
black hole through an external scalar gradient. Referencing this,
\cite{Horbatsch2011} has used observations of a black hole binary to  
constrain the cosmological time dependence for extremely light scalar 
fields.   The numerical calculations presented in \cite{Berti2013} also 
support this idea by showing that black holes moving through a 
non-uniform scalar gradient can emit scalar monopole and dipole radiation,
and \cite{Vincent:2016sjq,Ni:2016rhz} explore other possible 
observational effects of scalar hair.

In a previous paper, \cite{Davis:2014tea},
we made a preliminary investigation using an artificial matter 
distribution around a black hole -- an `accretion' thick sphere that
extended from $r=6GM$ out to large $r$ in a Schwarzschild black hole.
The purpose of that investigation was to first establish, within the rules of
the no hair theorems, that a black hole could indeed support scalar
hair. Next, analytic modelling of the scalar profile was undertaken
and compared in detail with numerical solutions so that we could
confidently make an estimate of the magnitude of the scalar profile 
for a generic scalar model. Finally, using this data, we explored
and estimated observational effects. Our aim here is to revisit the
crude (and unrealistic) model for the accretion sphere of the black
hole, and to use a more realistic disc model, and explore to
what extent the results we derived previously were dependent
on the assumed matter distribution around the black hole.

This paper is organised as follows: we briefly review screened 
modified gravity in \textsection\ref{screenedMG}. Next we solve 
for the scalar profile around a Schwarzschild black hole with an
accretion disc in \textsection\ref{Sch} followed by the Kerr black hole in 
\textsection\ref{kerr} and conclude by commenting on possible implications in 
\textsection\ref{astro}.

\section{Screened modified gravity}\label{screenedMG}

The way we solve the scalar equation of motion requires the scalar 
to evolve slowly (to be made more precise later) in response to the 
non-uniform matter density of an accretion disc. While several 
screening mechanisms exist, we focus on the frameworks 
where the additional scalar degree of freedom is constrained 
to have a large Compton wavelength compared to the length scale 
of astrophysical black holes.  As we argued in \cite{Davis:2014tea}, 
this is a common feature in several of the most popular screening 
mechanisms including the chameleon, the environmentally dependent 
dilaton, and the symmetron. 

The basic idea of screening is that the scalar mass or the coupling 
to matter (or both) is dependent on the local energy density, hence
in a dense environment such as our solar system, the field becomes 
`heavy', effectively decouples, and thus no fifth force modifications 
of gravity are present in such environments. On the other hand, at
cosmological scales and densities, the field is light and can give
rise to modifications of the gravitational interaction.

The relevant models of screened modified gravity include: 
the chameleon mechanism, \cite{Khoury:2003rn,Brax:2004qh},
which occurs when the mass of the scalar field, $m(\phi_0)$, 
is large enough to suppress the range of the scalar force;
the environmental dilaton, \cite{Brax:2010gi}, where the coupling
function between the scalar and matter fields and the mass alter in 
dense regions; and the symmetron, \cite{Hinterbichler:2010es}, 
where the coupling function switches off in dense environments. 
These mechanisms can be modelled generically with the Einstein frame action
\be
\label{eframe_general}
S=\int d^4 x \sqrt{-g}\left[-\frac{M_{p}^2}{2}R 
+\frac{1}{2}g^{\mu\nu}\partial_{\mu} \phi \partial_{\nu}\phi 
-V(\phi)\right]+S_m\left[\Psi_i,A^2(\phi)g_{\mu\nu}\right].
\ee
Where $M_{p}^2 = 1/8\pi G$ gives the Planck mass, $S_m$ represents the
matter action (denoted generically as $\Psi_i$), and $A(\phi)$ is the
conformal coupling between the Einstein and Jordan frames
$\tilde{g}_{\mu\nu} = A^2(\phi)g_{\mu\nu}$.  
The details of a particular theory are completely specified 
by the scalar potential $V(\phi)$ and the coupling function $A(\phi)$. 

Using this set-up, and identifying a conserved density 
$\rho \equiv -A^{-1}T_m$ in the Einstein frame \cite{Brax2012a}, 
the scalar equation of motion becomes
\begin{equation}\label{genphieq}
\square\phi = \frac{\partial}{\partial \phi}\left[V(\phi) +  (A(\phi)-1)\rho\right]\equiv\frac
{\partial  V_{\text{eff}}(\phi,\rho)}{\partial\phi}.  
\end{equation} 
Thus we see explicitly the density-dependent effective potential 
$V_{\text{eff}}(\phi,\rho)$ that is the source of the screening behaviour for 
chameleons, environmentally dependent dilatons, and symmetrons.  

\section{Schwarschild Black Hole}\label{Sch}
 
We start by exploring the accretion disc around a Schwarzschild black hole,
as a warm up for the full Kerr problem. In order to proceed, we 
require a simple model for the physical set-up. 
We assume that we have some background ambient density field, ignoring 
for now any general isotropic build up of matter in the neighbourhood of the 
black hole. Superposed on this ambient density field is the accretion disc, 
which is generally highly concentrated in the equatorial plane, extending
out from the Innermost Stable Circular Orbit (ISCO) of the black hole.

We model the accretion disc by a uniform density $\delta-$function on the 
equatorial plane extending from the ISCO to some (arbitrary) outer radius 
$r_1$ characteristic of the accretion disc or the galactic plane. 
This has the desirable property of being disc-like 
and constrained in a 2-dimensional plane around the black hole,
although the constant density profile is an idealisation.
Astrophysically realistic accretion disc models involve complex 
fluid dynamics typically requiring numerical modelling
(see \cite{Abramowicz:2011xu} for a detailed review) and are beyond 
the scope of this work. However our results should 
capture the salient features of these more involved models, 
have the particular benefit of being amenable to analytic analysis, 
and should provide a reasonable estimate for the scalar field profiles. 

The idea is to analyse the scalar equation \eqref{genphieq}
in the strongly curved geometry near the black hole event horizon for this
idealised disc source. Our aim is to proceed as far as possible analytically,
so that we can obtain general results and features of the solution that can
be used in a wider range of models than if we were to pick specific
potentials, couplings, and solve the problem numerically.

The first step is to choose an appropriate coordinate system for
the analysis. As we will see, it turns out to be most rewarding to 
rewrite the Schwarzschild metric in ``Weyl'' form, where the radial
and polar angles $\{r,\theta\}$ are re-badged as a pair of cartesian-like
coordinates $\{x,y\}$ such that the $x-y$ part of the metric is
conformally flat:
\begin{align}
ds^2 = e^{2\lambda}dt^2 - e^{2(\nu-\lambda)}\left(dx^2+dy^2 \right)-\alpha^2e^
{-2\lambda}d\varphi^2
\end{align}
This is the general Weyl metric form, but for the Schwarzschild solution,
the functions take the form
\be
\alpha \equiv x \;,\qquad
\lambda = {\frac12} \ln {X_+ - Y_+ \over X_- - Y_-}\;,
\qquad
\nu = {1\over2} \ln {(X_+X_- + Y_+ Y_- + x^2) \over 2X_+X_-}
\ee
with
\be
Y_\pm = y \pm GM \qquad , \qquad X_\pm^2 = x^2 + Y_\pm^2
\ee
We can return to the familiar $\{r,\theta\}$ Schwarzschild coordinates
via the transformation 
\be
x^2 = r(r-2GM) \sin^2\theta \;\; , \;\;\; y=(r-GM)\cos\theta
\label{w2sch}
\ee

Now consider the scalar field equation in this background. Our model
for the accretion disc supposes that while it may have highly nontrivial
local dynamics, these average out to an approximately uniform density
profile strongly localised in the equatorial plane. The scalar profile therefore
will be dependent essentially on only the radial and polar coordinates.
This is the reason for choosing this less well known Weyl coordinate system 
for the Schwarzschild metric -- the wave operator turns out to have a 
simple form if the scalar depends only on $x$ and $y$,
being proportional to a flat-space cylindrical Laplacian, leading to 
the equation of motion for $\phi$:
\be
\Box \phi \equiv \frac{1}{\sqrt{-g}}\partial_{\mu}
\left(\sqrt{-g} g^{\mu\nu}\partial_{\nu} \right) \phi 
= e^{-2(\nu-\lambda)} \left [
\frac1x \frac{\partial~}{\partial x} \left ( x \frac{\partial\phi}{\partial x} \right)
+ \frac{\partial^2\phi}{\partial y^2} \right] 
= \frac{\partial V_{\text{eff}}(\phi, \rho)}{\partial\phi}
\label{phiwaveeq}
\ee
for the appropriate $V_{\text{eff}}$. 

In \cite{Davis:2014tea}, we used a
much simpler background matter density in the vicinity of the black
hole, and found approximate analytical solutions, comparing them to
full numerical solutions for the scalar profile. With the exception of 
long Compton wavelength symmetrons, these were largely similar, 
with screened scalars having broadly similar profiles that peaked 
near the event horizon. 
We will therefore consider the Chameleon model in this
paper, and take a large Compton wavelength compared to the typical black hole 
length scales. The current experimentally constrained model parameters of 
environmentally dependent dilatons and symmetrons 
also put them in the same category of long Compton wavelength scalars. 

In the chameleon models, it is typically assumed that the 
coupling function is to a good approximation an exponential:
\be 
A(\phi) = e^{\beta\phi/M_p},
\ee
where $\beta$ is nearly constant over the range of field values of interest.
In addition, a typical chameleon potential is usually taken to be 
\be\label{chV1}
V(\phi) = M^{4+n} \phi^{-n}=  V_0 \phi^{-n},
\ee
where $n\geq 1$ is an integer of order one, and we define $V_0\equiv M^{4+n}$ to 
simplify notation. Keeping only the leading order term from the coupling function, we 
see that the effective potential is
\be\label{veff}
\Veff(\phi,\rho)\approx \frac{V_0}{\phi^{n}} + \frac{\rho\beta\phi}{M_p},
\ee
minimised at
\be
\phi_{\text{min}}^{n+1} =\frac{nV_0M_p}{\rho\beta}.
\ee
The mass of small fluctuations of the field around this minimum is
\be\beal
m^2(\rho) &=\left.\Veff(\phi,\rho),_{\phi\phi}\right|_{\phi_{\text{min}}} \\
&\approx \frac{\rho\beta}{M_p} \left[
(n+1)\left(\frac{\rho\beta}{nV_0M_p}\right)^{\frac{1}{n+1}} 
+\frac{\beta}{M_p}\right]\label{ch_masseq}
\eeal\ee
which, as required, increases monotonically with $\rho$.

To find $\rho$, note that in the Weyl coordinates the equatorial 
plane corresponds to $y=0$, and the ISCO radius, $r=6GM$ for the
Schwarzschild black hole, corresponds to $x_0 = 
2\sqrt{6} GM_{BH} = \sqrt{6} r_s$, where $r_s=2GM_{BH}$ is the
Schwarzschild radius.
Thus the accretion disc model we are using has the density profile
\be 
\rho \rightarrow \rho_0 + \rho_1 \delta(y)\Theta[x-x_0]\Theta[x_1-x] 
\label{discprofile}
\ee
in Weyl coordinates.

We now make two simplifying assumptions in order to explore
scalar solutions analytically. Firstly, we use this crude model
for the disc \eqref{discprofile}. Secondly, we assume that the
solution for the scalar is dominated by the effect of $\rho_1$, the
accretion disc itself; essentially this means we expand our scalar 
around $\phi_{\text{min}}$
\be
\phi \sim \phi_{\text{min}} + \delta\phi
\ee
where $\phi_{\text{min}}$ is the background scalar field profile due
to the ambient background density $\phi_0$.
Finally, we assume that the mass of the scalar is negligible.
This will be a good approximation within the Compton radius of the
scalar, and provided our system does not extend over many
Compton wavelengths, should give a realistic picture for the scalar
profile.

Making these assumptions, \eqref{phiwaveeq} reduces to a 
Poisson equation for $\delta \phi$:
\be
(x \delta \phi_{,x})_{,x} + x\delta\phi_{,yy}  = 
\frac{\beta\rho_1(\mathbf{r})}{M_p} \,
x e^{2(\nu-\lambda)}
\label{laplacephi}
\ee
for which we can use the massless scalar Green's
function to obtain:
\be 
\delta\phi = -\frac{\beta}{4\pi M_p}\int d^3 \mathbf{r'} \frac{\rho_1(\mathbf{r'})
e^{2(\nu-\lambda)} }{\lvert \bold{r} - \bold{r^{'}} \lvert}
\ee

The accretion disc model \eqref{discprofile} localises this integral to the 
equatorial plane where
\be
e^{2(\nu-\lambda)}\mid_{y'=0} = 
\frac{(\sqrt{4x^{\prime2}+r_s^2}+r_s)^2}{4x^{\prime2}+r_s^2}
\ee
and using
\be
\int^{\pi}_{-\pi}\frac{d\varphi'}{\sqrt{y^2+x^2+x^{\prime2}-2xx'\cos(\varphi-\varphi')}} 
= \frac{4}{\sqrt{(x+x')^2+y^2}}\; \text{K}\left[\frac{4xx'}{(x+x')^2+y^2} \right]
\ee
where K is the complete elliptic integral of the first kind,
$\delta\phi$ becomes 
\be
\delta\phi = -\frac{\beta \rho_1r_s}{\pi M_p}\int^{x_1}_{x_0}
\frac{(\sqrt{4x^{\prime2}+r_s^2}+r_s)^2}{x^{\prime2}+r_s^2}
\;\text{K}\left[\frac{4xx'}{(x+x')^2+y^2} \right]\;
\frac{x' dx'} {\sqrt{y^2+(x+x')^2}}
\label{deltaphiintegral}
\ee

Once we specify an $x_1$, we can integrate up this expression to obtain
$\delta\phi$, which we will do presently. However, for the moment we would
like to obtain an order of magnitude estimate for $\delta\phi$, and its dependence
on the various parameters analytically. First, extract the dependence on the
black hole mass by rescaling $\hat{x} = x/GM_{BH} = 2x/r_s$:
\be
\delta\phi 
= -\frac{\beta \rho_1 r_s^2}{4 M_p}\; {\hat{\delta\phi}}
= -\frac{\beta \rho_1 r_s^2}{4\pi M_p}\;
{\cal{I}} [\hat{x},\hat{y}]
\label{deltaphi}
\ee
where
\be
{\cal{I}} [\hat{x},\hat{y}]= \int^{\hat{x}_1}_{2\sqrt{6}}
\left(\frac{(\sqrt{\hat{x}^{\prime2}+1}+1)^2}{\hat{x}^{\prime2}+1}\right)
\;\text{K}\left[\frac{4\hat{x}\hat{x}{'}}{(\hat{x}+\hat{x}')^2+\hat{y}^2} \right]\;
\frac{\hat{x}' d\hat{x}'} {\sqrt{\hat{y}^2+(\hat{x}+\hat{x}')^2}}
\label{schintegral}
\ee
The prefactor in \eqref{deltaphiintegral} gives the parameter dependence for
the scalar, and we now approximate \eqref{schintegral} to get an estimate
of the order of magnitude of $\hat{\delta\phi}$.
\begin{figure}[!htb]
\centering
\includegraphics[scale=0.7]{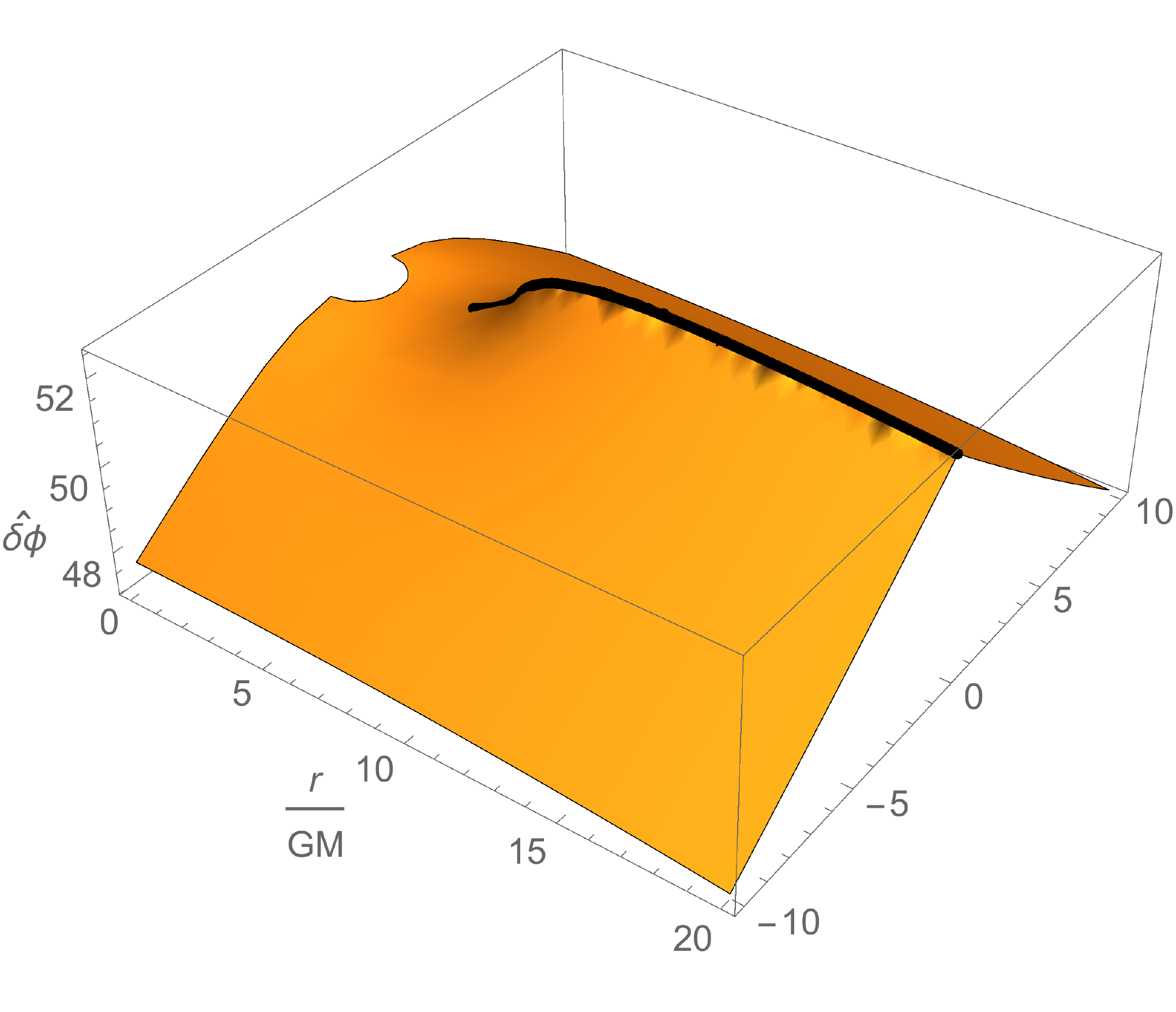}
\caption{3D plot illustrating the (normalised) scalar field profile $\hat{\delta\phi}$
around the Schwarzschild accretion disc.  
The start of the accretion disc (shown as a thick black line)
can be clearly seen around $x\approx5$.} 
\label{fig:schprofile}
\end{figure}

The first term for ${\cal I}$ in brackets is monotonically decreasing, and for 
$\hat{x}'\geq2\sqrt{6}$ lies in the range $[1,36/25]$, thus we approximate
this term by $1$. The elliptic function is roughly $\pi/2$ unless $\hat{y}\approx0$,
$\hat{x} \approx \hat{x}'$, i.e., on the accretion disc, however since the 
singularity of K is logarithmic, this will not give a huge enhancement to
the integral, thus we approximate this contribution to the integrand by
$\pi/2$. This leaves us with the final term, that can be integrated exactly
to give
\be
{\cal I} [\hat{x},\hat{y}] \simeq \frac\pi2 \left [
R_1-R_0 - \hat{x} \ln \frac{\hat{x} + \hat{x}_1 + R_1}
{\hat{x} + \hat{x}_0 + R_0}\right]
\label{schintform}
\ee
where we have written $R_i = \sqrt{(\hat{x}+\hat{x}_i)^2+y^2}$ for clarity. 
Expanding this at large $\hat{r}=\sqrt{\hat{x}^2+\hat{y}^2}$ gives
${\cal I} \approx \pi (\hat{x}_1^2-\hat{x}_0^2)/2\hat{r}$, i.e.\ the expected 
``$1/r$'' fall-off of a massless field. Near the black hole and disc, $\hat{x} \approx
\hat{x}_0, \hat{y} \approx 0$, and ${\cal I} \approx \pi \hat{x}_1/2$. 
We therefore obtain an order of 
magnitude estimate for the magnitude of the chameleon near the disc of:
\be
\delta\phi \approx -\frac{\beta \rho_1 r_s^2}{8 M_p}\; 
\frac{r_1}{r_s}
\label{deltaphihor}
\ee
Note that this will be an underestimate, since in each case
in the integrand, our estimate was the lower, though more consistent,
value of the function. At large distances from the disc, the profile
becomes very accurate, but closer in, we may expect some discrepancy.
\begin{figure}[!htb]
\centering
\includegraphics[scale=0.7]{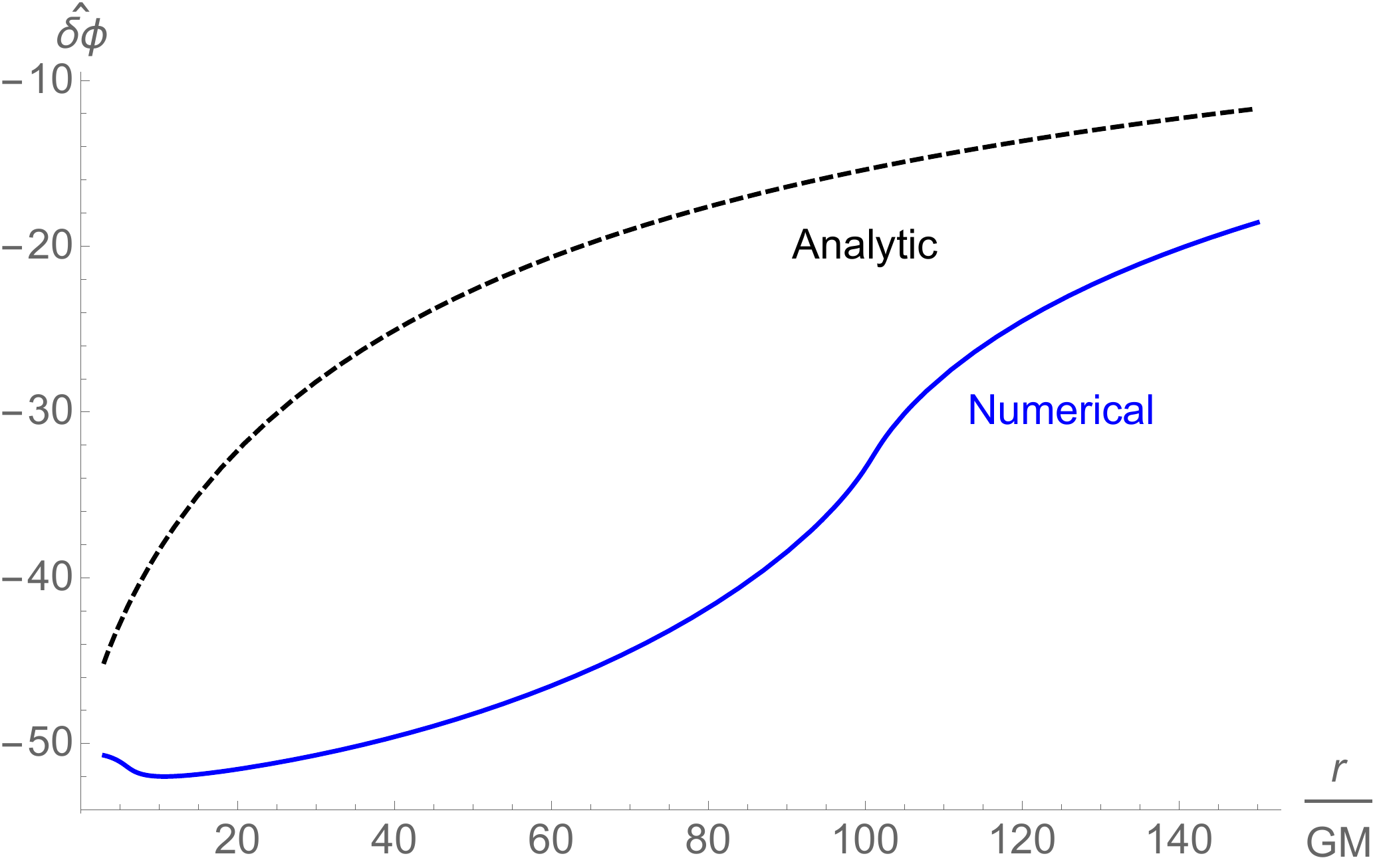}
\caption{A comparison of the analytic (under)estimate and the numerically
integrated result plotted in Schwarzschild `cartesians', $(r\sin\theta,r\cos\theta)$,
at constant $r\cos\theta = GM$. The largest disparity between the two curves
is along the length of the accretion disc ($6GM<r<100GM$), with the
curves approaching at larger $r$. The peak estimate for $\delta\phi$ agrees
within $10\%$ however.} 
\label{fig:comparison}
\end{figure}

In order to check this estimate, we integrated \eqref{schintegral}
using mathematica; figure \ref{fig:schprofile} shows the scalar field 
plotted in Schwarzschild coordinates $(r\sin\theta/GM, r\cos\theta/GM)$,
and figure \ref{fig:comparison} shows the accuracy (or otherwise) of this
estimate. The presence of the accretion disc clearly causes the 
scalar field to respond and lifts it from its ambient background value.
The disc itself is evident in the plot from the sharp crease in the profile, 
resulting from the integrated singularity of the elliptic function. This
is clearly an artefact of the fact we have modelled the disc with a hard
$\delta-$function profile. In a more realistic scenario, the accretion disc 
while strongly localised near $y=0$, will have some spread on either side,
and we would expect this kink discontinuity to smooth out. 
\begin{figure}[!htb]
\centering
\includegraphics[scale=0.5]{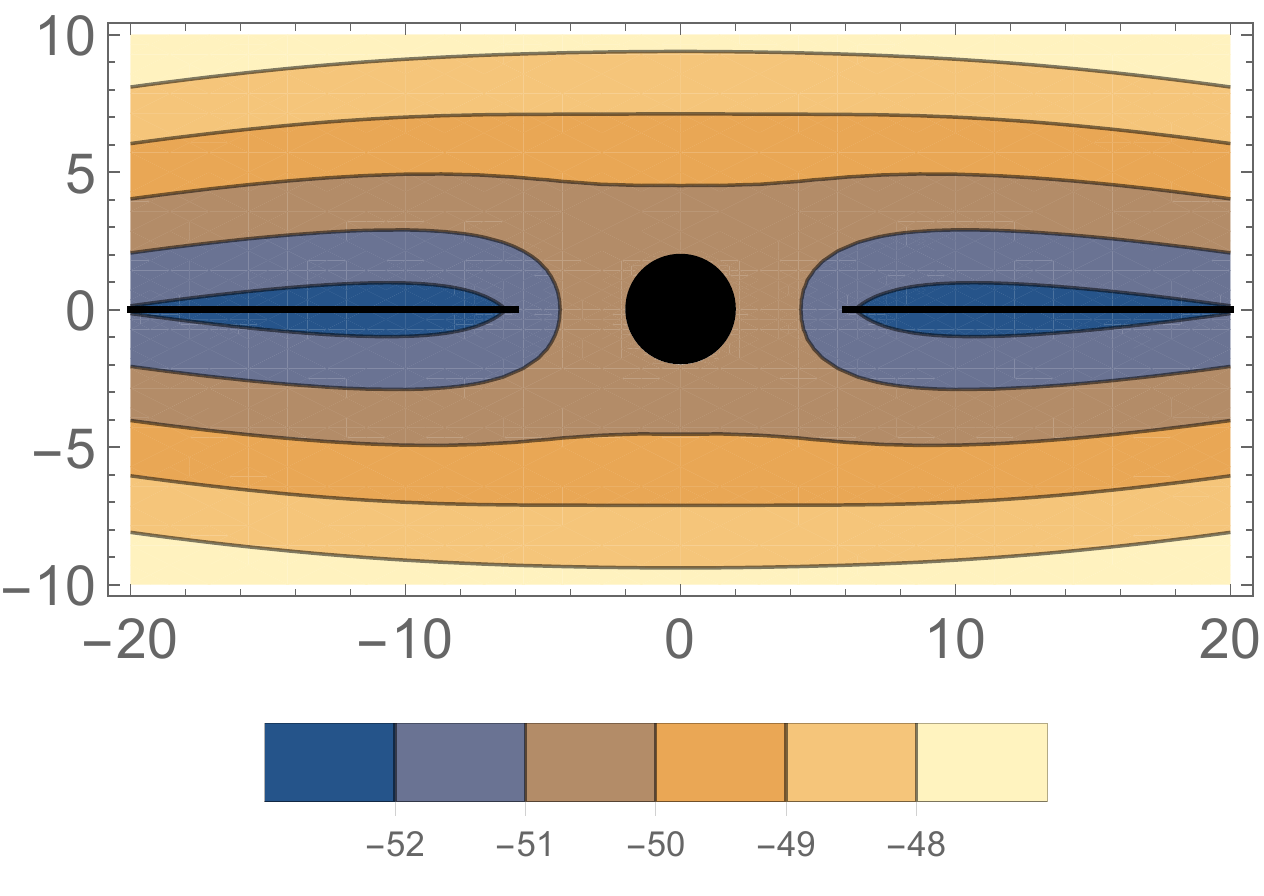}~
\includegraphics[scale=0.5]{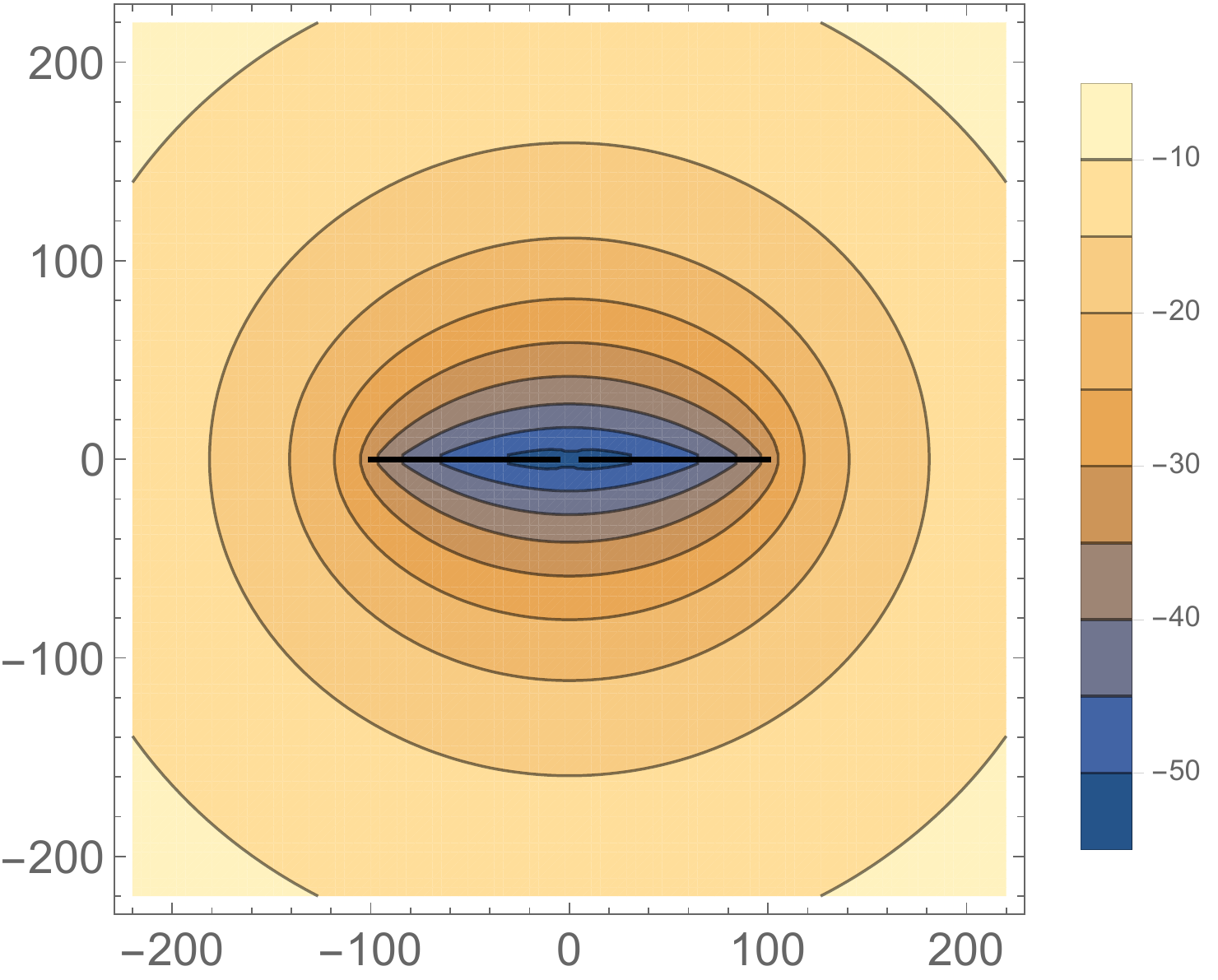}
\caption{Comparison of near field and far field fall off around the 
Schwarzschild accretion disc, indicated in each plot by the horizontal 
thick black line.} 
\label{fig:phiprofile}
\end{figure}

On large distances, the scalar rolls back to its ambient value, as expected
from the physics and analytic approximation. Figure \ref{fig:phiprofile}
shows the near and further field solutions for $\delta\phi$. In practise,
once the scalar Compton wavelength scale is reached, the field will
then transition to the typical exponential fall-off expected of the massive
field profile.

\begin{figure}[!htb]
\centering
\includegraphics[scale=0.6]{schw.pdf}
\includegraphics[scale=0.6]{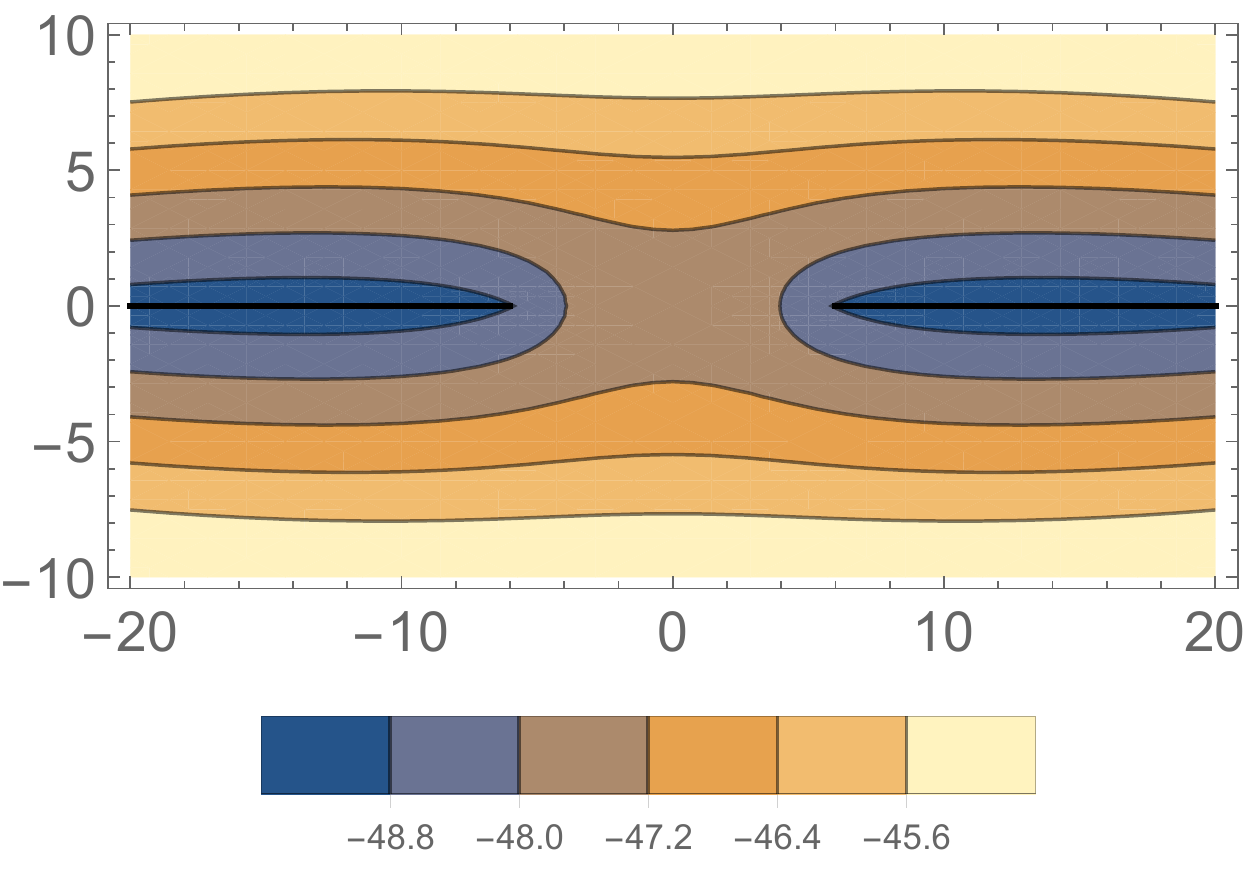}
\caption{Comparison of scalar profile with and without a black hole. The
disc profile is taken to have the same form, starting at $r=6GM$, and
ending at $r=100GM$.} 
\label{fig:nobh}
\end{figure}

It is interesting to query to what extent the scalar profile is due to
the geometry of the black hole, and what to the matter distribution,
which would in any case cause the scalar to shift. Figure \ref{fig:nobh}
shows a comparison of the scalar perturbation from the disc plus
black hole, to the disc only. This clearly shows that primary feature
of the scalar being pulled from its equilibrium value by the dense matter
is due to the disc matter density, however, the black hole does
impact on the magnitude of the effect (if one looks at the contour
values) increasing it by about $10\%$. On the one hand, this
might suggest that the black hole is not that relevant, however, 
the disc would obviously not be there without the black 
hole to drive it. In addition this confirms the fact that in spite of the
strong gravity regime of the black hole, and the notion that the
event horizon is somehow ``special'', the scalar still behaves and
responds to its environment, with the black hole providing a
marginal boost to the local matter environment effects. As a result,
it seems counter-intuitive that a black hole would behave differently
towards a scalar than the local galactic medium, as suggested for
example for Vainstein screening \cite{Hui:2012jb}.

\section{Kerr Geometry}\label{kerr}

Having discussed the static, spherically symmetric case it is now surprisingly
straightforward to turn to the more physically interesting case of the rotating 
Kerr geometry, usually written in spherical polar Boyer-Lindquist 
coordinates as
\be \label{kerrmetric}
ds^2 = \frac{\Delta-a^2\sin^2\theta}{\Sigma}dt^2 
+ \frac{4GMar\sin^2\theta}{\Sigma}dtd\varphi
- \frac{\beta}{\Sigma}\sin^2\theta d\varphi^2 
- \frac{\Sigma}{\Delta}dr^2-\Sigma d\theta^2 
\ee
where $a=J/M$ and 
\be
\beal
\Sigma &= r^2 + a^2\cos^2\theta \\
\Delta &= r^2 - 2GMr+a^2\\
\Gamma &= (r^2+a^2)^2 - \Delta a^2\sin^2\theta
\eeal
\ee

Following the method described for the simpler Schwarzschild geometry,
we begin by rewriting the metric in Weyl coordinates \cite{Gregory:2013xca}
\be\label{kerrweyl}
ds^2 = e^{2\lambda}dt^2 - \alpha^2 e^{-2\lambda}[d\varphi+Bdt]^2 
- e^{2(\nu-\lambda)}(dx^2+dy^2)
\ee
where 
\be
x\equiv\alpha = \sqrt{\Delta}\sin\theta\;\;,\quad
y = (r-GM)\cos\theta
\ee
To get the Weyl functions, we first define
\be
\beal
Y_{\pm} &= y \pm \sqrt{G^2M^2 - a^2},&&\\
X^2_{\pm} &= x^2+y^2_{\pm} \qquad 
\Rightarrow \quad X_{\pm} &&= r-GM\pm\sqrt{G^2M^2-a^2}\cos\theta
\eeal
\ee
Giving
\be
e^{2\lambda} = \frac{\Delta\Sigma}{\Gamma}\;\;,\quad
B = \frac{2aGMr}{\Gamma}\;\;,\quad
e^{2(\nu-\lambda)} = \frac{\Sigma}{X_+X_-}
\ee 
where
\be
r = \frac{X_+ + X_-}{2} + GM\;\;,\quad
\cos\theta = \frac{X_+-X_-}{2\sqrt{G^2M^2-a^2}}
\ee

Once again, we model the accretion disc by the simplified energy
distribution \eqref{discprofile}, and insert in \eqref{laplacephi} now
the Kerr measure
\be\label{kerrmeasure}
e^{2(\nu-\lambda)} \mid_{y=0} = \frac{(\sqrt{\hat{x}^{\prime2}+1-
\hat{a}^2}+1)^2}{\hat{x}^{\prime2}+1 - \hat{a}^2}
\ee
where, as before, we have rescaled our Weyl coordinates,
and $\hat{a} = a/GM_{BH} \in [0,1]$.

It is easy to see that the scalar equation of motion remains mostly 
unaffected by the addition of rotation into the geometry. Its 
functional form is unchanged in Weyl coordinates, although the 
multiplicative factor of $e^{2(\lambda-\nu)}$ must now take the Kerr
form. The general expression \eqref{deltaphiintegral} therefore
remains the same, with $r_s = 2GM_{BH}$ representing now the
black hole mass rather than the horizon radius, and with the
integral function replaced by the appropriately modified
Kerr expression:
\be
{\cal{I}}_{\text{Kerr}} [\hat{x}_1]= \int^{\hat{x}_1}_{\hat{x}_0}
\left(\frac{(\sqrt{\hat{x}^{\prime2}+1-\hat{a}^2}+1)^2}
{\hat{x}^{\prime2}+1-\hat{a}^2}\right)
\;\text{K}\left[\frac{4\hat{x}\hat{x}{'}}{(\hat{x}+\hat{x}')^2+\hat{y}^2} \right]\;
\frac{\hat{x}' d\hat{x}'} {\sqrt{\hat{y}^2+(\hat{x}+\hat{x}')^2}}
\label{kerrintegral}
\ee
where $\hat{x}_0$ is the rescaled ISCO value of $x$. The general expression
for $\hat{x}_0$ in terms of $\hat{a}$ is somewhat unwieldy, however, the
key feature is that $\hat{x}_0$ decreases as $\hat{a}$ increases, eventually
merging with the event horizon at $\hat{x}_0=0$. 

\begin{figure}[h]
\centering
\includegraphics[scale=0.5]{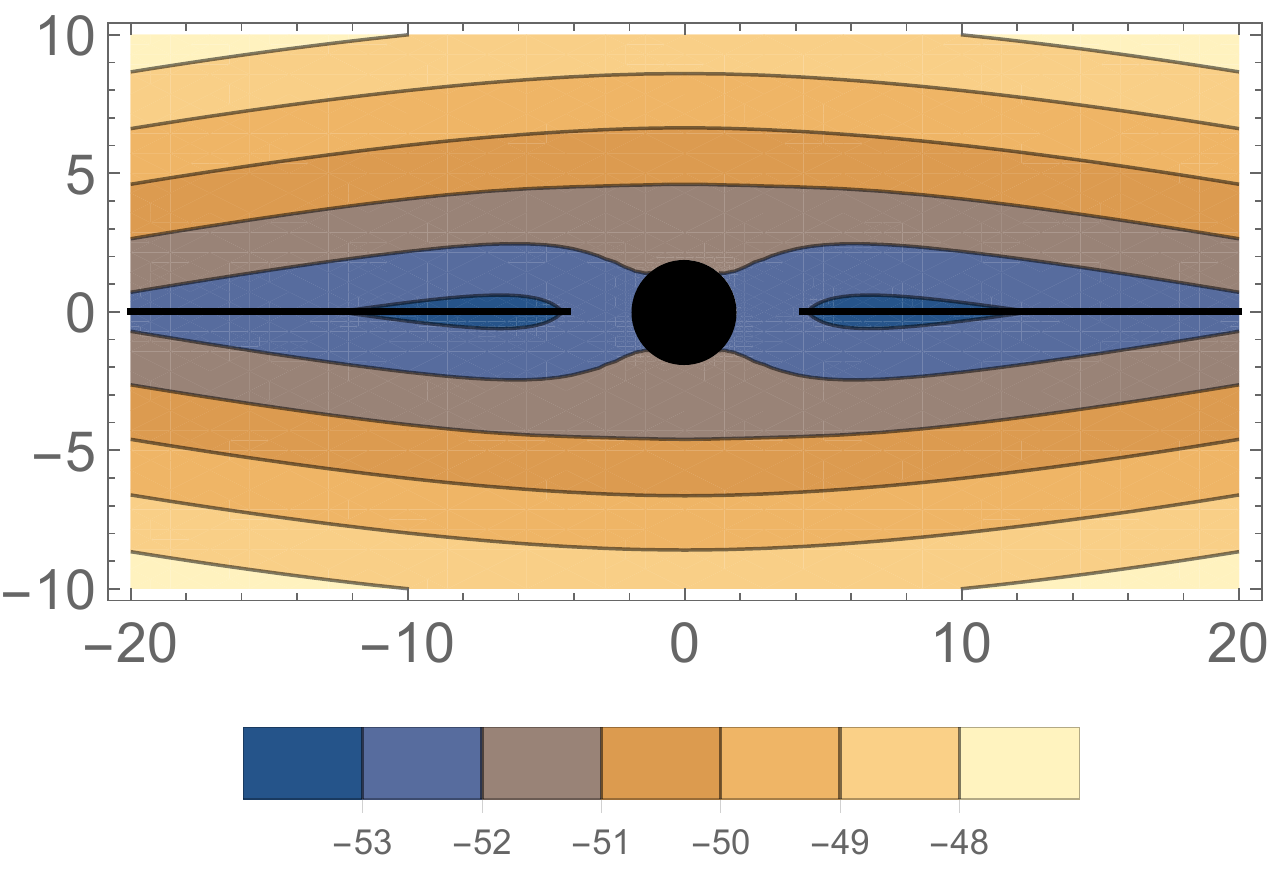}~
\includegraphics[scale=0.5]{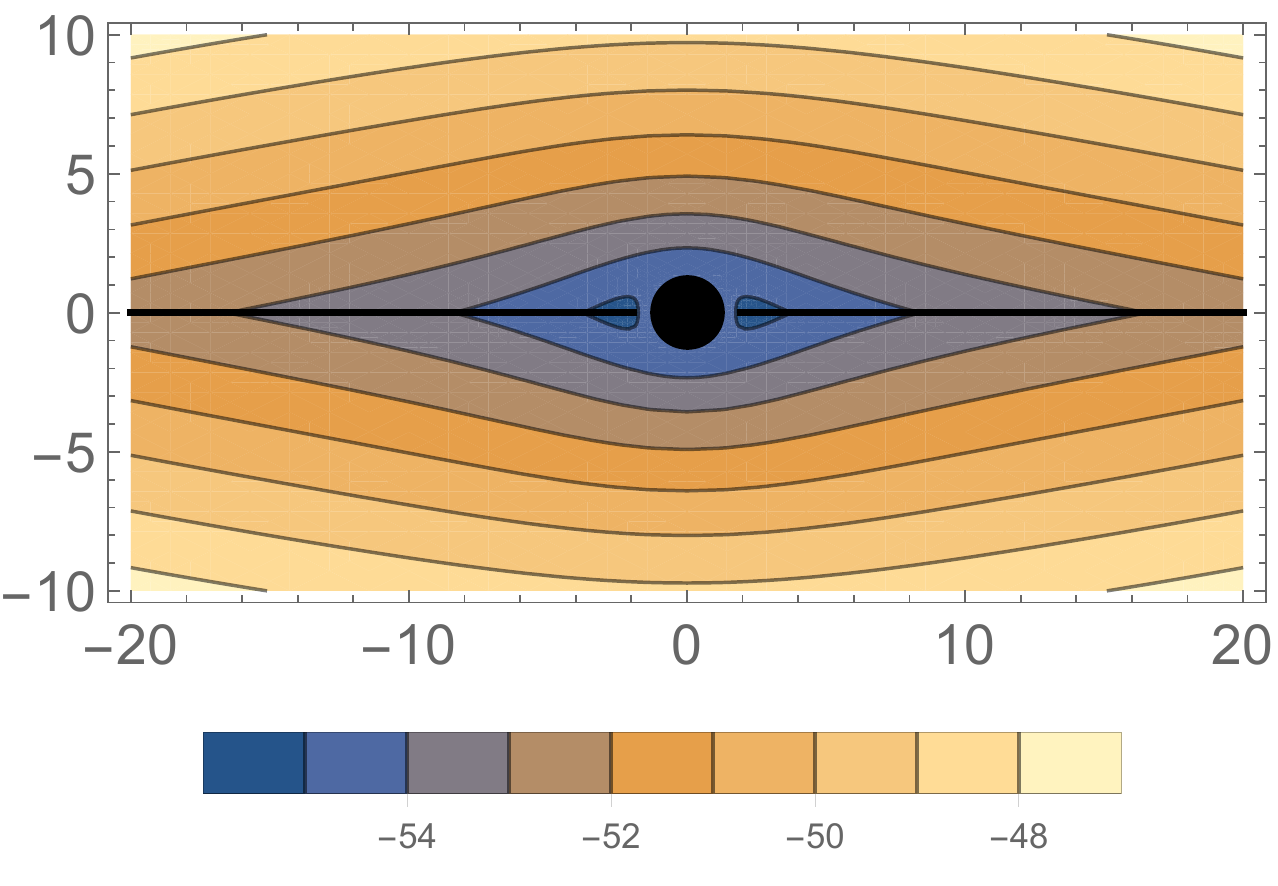}
\caption{The normalized scalar profile $\hat{\delta\phi}$ 
plotted in cartesian coordinates for increasing rotation: $a/GM=0.5, 0.95$. 
The accretion disc is represented as a solid black line and the black hole 
as a black circle in the middle.}
\label{fig:sch}
\end{figure}

As before, the elliptic integral contributes roughly a constant, except
very near the accretion disc where is gives a slight uplift to the integral.
The final term is unchanged, however, the first factor, coming from
the $e^{2(\nu-\lambda)}$ term, is now potentially rather different if
the black hole is at, or very near, extremality. For $\hat{a}\sim1$,
this term is roughly $1/\hat{x}^{\prime2}$, and thus the integral 
generically diverges logarithmically as $\hat{x}_{\text{ISCO}}\to0$ 
in the extremal limit, with a linear divergence at $\hat{x}=\hat{y}=0$.
Although this sounds alarming, because of the precipitous drop in 
the ISCO as $a \to a_{\text{ext}}$, $x_{\text{ISCO}}\sim 2^{\frac23}
(1-\hat{a})^{\frac13}$, this only contributes an uplift to ${\cal{I}}_{\text{Kerr}}$ 
of order a few for any realistic astrophysical black hole. We therefore
expect a very similar expression to \eqref{schintform} for an
estimate of the integral. The primary difference will be that the
strongest shift of the scalar will be near the ISCO, which will
be much closer to the event horizon of the black hole, therefore
correspondingly a sharper profile. This is borne out by the 
numerical integrations shown in figure \ref{fig:sch}, which
show the scalar profile in Boyer-Lindquist coordinates 
for $\hat{a} = 0.5$ and $0.95$. 

Our overall conclusion therefore is that the disc pulls the scalar from
its ambient value to a central order of magnitude of \eqref{deltaphihor}.
Rotating black holes give a stronger effect,
but only by about $5-10\%$, even for a nearly extreme black hole.

\section{Summary and Discussion}\label{astro}

In the main part of this paper, we presented an analytic analysis of the scalar
profile around a Schwarzschild or Kerr black hole with an accretion disc. With
the assumption of a very sharp profile accretion disc, modelled by a $\delta-$function,
the scalar depends only on the radial distance and angle from the rotation axis.
Using the less-well known Weyl co-ordinate system for the black hole geometry,
the scalar equation of motion simplifies considerably to a form for which a Green's
function is known. We used this to estimate analytically the scalar field profile,
then confirmed this by a simple numerical integration.

The scalar has a nontrivial profile around the black hole, and is `pinned' to its
largest value on the accretion disc, very near the ISCO. The main result is the
scalar displacement amplitude \eqref{deltaphihor}:
\be
\delta\phi = -\frac{\beta \rho_1 r_s^2}{8 M_p}\;
{\cal{I}}
\label{gendeltaphi}
\ee
where ${\cal I} = r_1/r_s$ represents the extent of the dense accretion flow.
We now turn to a discussion of the potential astrophysical consequences 
and observable effects of this scalar profile.
Our initial assumptions about the accretion disc being a small addition,
so as to not disrupt the background Schwarzschild or Kerr geometry appreciably, 
would guarantee the magnitude of any effect due to the scalar field to be small. 
 
The most obvious effect to consider would be an additional fifth force felt 
by any test particle in the vicinity of the black hole accretion disc system. 
Though it is entirely possible that these additional forces would cause the 
structure of the accretion disc to be non-trivially modified, such effects 
would require astrophysical modelling beyond the scope of the present work. 
We considered this possibility in \cite{Davis:2014tea}, where we concluded
that the effect would be too small to be observed for a coupling $\beta$ of 
$O(1)$ and while the magnitude of the scalar is similar here since the modelling
of the accretion disk is rather different we should be able to get a better
estimate of the fifth force.

The effects of the scalar gradient on a test particle can 
easily be estimated as the ratio of the fifth force to the Newtonian force is 
\be\label{forcecomp}
\frac{|F_{\phi}|}{|F_N|}\approx \left(\frac{r}{r_s}\right)^2 
\beta(\phi)|\vec{\nabla}\phi|\frac{M_{BH}}{M_p^3}.
\ee
\eqref{gendeltaphi} allows us to approximate this as 
\be
10^{-11}\beta^2 {\cal{I}}\lesssim \frac{|F_{\phi}|}{|F_N|}\lesssim 10^{-3}\beta^2 {\cal{I}}
\ee
where we use $\rho_1 \sim 10^{21} \rho_{cos}$, 
$\vec{\nabla}\phi \sim \delta\phi / R_1$ and 
$M_{BH} \sim 10^6-10^{10} M_{\odot} \sim 10^{44} - 10^{48} M_p$
We note that while this ratio is small, for $\beta \sim \mathcal{O}(10-10^2)$, 
which is an allowed parameter range for the chameleon model
\cite{Elder:2016yxm}, the effects 
could be an appreciable percentage of the Newtonian force in some cases. 

It is obvious from the force estimation that to evaluate the relevance of our 
scalar field profile in the accretion disk, we should compare the emission
of any scalar radiation to gravitational radiation.
We can do this if we view our static black hole model as the supermassive 
partner of an extreme mass ratio inspiral (EMRI) binary system. Such EMRI 
systems typically consist of a stellar mass compact object orbiting a 
supermassive black hole and in GR they emit gravitational radiation 
at a rate approximated to leading order in $\dot r$ by the quadrupole formula. 
Comparing this to the scalar radiation gives, 
\be
\left|\frac{\dot{\mathcal{E}}_{\phi}}{\dot{\mathcal{E}}_{GR} }\right| 
\sim \beta\left(\frac{R_0}{R_s}\right)^{\frac{9}{2}}
\left(\frac{\delta\phi}{\delta r}\right)
\frac{M_{BH}}{M_p^3}\left[\frac{M_{BH}}{m_t}\right].
\ee
Where $m_t$ is a test mass. For $m_t \sim M_{\odot}$ and
\be
10^6 \leq \frac{M_{BH}}{M_{\odot}} \leq 10^{10}
\ee
we get, 
\be
\beta^2 10^{-5}\lesssim \left|\frac{\dot{\mathcal{E}}_{\phi}}
{\dot{\mathcal{E}}_{GR} }\right| \lesssim \beta^2 10^7
\ee

For ultramassive black holes the ratio of scalar to gravitational radiation is
very large. Since observations of such objects are currently rare we do not
know if the fate of stellar mass inspirals would be similar to those in 
supermassive black holes. It would be interesting to see if such scalar 
radiation emitted by inspirals could be observed. 

Another interesting effect which could potentially be observable is a shift 
in the atomic spectra. 
The scalar field in the chameleon model couples to matter and therefore 
the effective mass of elementary particles within the accretion disk will 
now receive a small correction proportional to $\delta \phi$  \cite{Brax:2010gp}. 
For the electron we have, 
\be
m_e(\phi) = m_e(1 + \frac{\delta\phi} {M_{p}})
\ee
This in turn should add a correction to all atomic spectra. 
In particular it will change the Balmer and Lyman $\alpha$ series via an 
effect on the Rydberg constant and as it turns out both of these are observable 
in the quasar spectra, see eg SDSS \cite{VandenBerk:2001hc}. 
The Rydberg constant, in natural units, may be expressed as, 
\be
R_{\infty} = \frac{\alpha^2 m_e}{4\pi} 
\ee
where $\alpha$ is the fine structure constant. The shift in the energy levels, 
assuming a correction to just the Rydberg constant, is then
\be
\frac{\Delta E}{E} = \frac{\delta\phi} {M_{p}} 
\ee
since $\delta\phi$ is negative this gives a negative shift in the energy levels of order
\be
\beta 10^{-13} < \left| \frac{\Delta E}{E} \right| < \beta 10^{-5}
\ee
The shift is the same for the $H_{\alpha}$, $H_{\beta}$ and $Ly_{\alpha}$ lines. 
It remains to be seen whether or not this can be detected with future surveys,
though for $\beta \sim \mathcal{O}(10)$ the shift could be appreciable for
supermassive black holes. 

If the scalar couples to photons via loop effects \cite{Brax:2010uq} then there 
will be a shift in hyperfine splitting as well. However, this is beyond the 
scope of this paper and requires further investigation. Such hyperfine splitting
would be easier to distinguish from the spectral lines required to determine
the quasar redshift.   

Our results are also strongly indicative of a breakdown of the no-hair theorems 
when applied to realistic astrophysical black holes. The presence of a small 
amount of matter surrounding the black hole is enough to violate the 
stringent conditions required for the no-hair theorems to be valid and 
as such their use in breaking the model degeneracy between various 
modified gravity scenarios would be very limited \footnote{The only case 
where no-hair theorems will provide a conclusive result is, in fact, in 
scenarios where they predict large deviations from GR in vacuum}. 

It has previously been argued that black holes and stars in the centre of 
galaxies would behave differently under the effect of the additional scalar 
force because, 
while the star will feel the effects of such a force, the black hole would be 
protected by a no-hair theorem \cite{Hui:2012jb,Hui:2012qt}. Our results 
show that this is not the case and that in any realistic 
astrophysical scenario the situation is likely to be much more complex and 
the effects of the accretion disk has to be taken into account when computing
the force experienced by an astrophysical black hole in the centre of a galaxy
compared to that experienced by the stars. This is beyond the scope of this
investigation. 

\acknowledgments

We would like to thank Raul Abramo and Andrei Frolov
for discussions. 
ACD would like to thank the Perimeter Institute, and RG the Aspen Center for
Physics, for hospitality while this work was being completed. ACD is supported
in part by STFC under grants ST/L000385/1 and ST/L0000636/1.
RG is supported in part by STFC (Consolidated Grant ST/J000426/1),
in part by the Wolfson Foundation and Royal Society, and in part
by Perimeter Institute for Theoretical Physics. 
Research at Perimeter Institute is supported by the Government of
Canada through Industry Canada and by the Province of Ontario through the
Ministry of Economic Development and Innovation. 
RJ is supported by the Cambridge Commonwealth Trust and 
Trinity College, Cambridge. 
This work was supported in part by the National Science Foundation under 
Grant No. PHY-1066293.
\providecommand{\href}[2]{#2}
\begingroup\raggedright

\end{document}